\documentclass[smallextended,natbib]{amsart}
\usepackage{graphicx}
\usepackage{amsmath}

\begin{document}

\title{The Game of Pure Strategy is solved!}

\author{Glenn C. Rhoads}

\address{Glenn C. Rhoads:
       University of Maryland, University College, 3501 University Blvd. East,
       Adelphi, MD 20783
}
\email{grhoads@faculty.umuc.edu}
\author{Laurent Bartholdi}
\address{
       Laurent Bartholdi:
       Bunsenstra\ss e 3-5, D-3703 G\"{o}ttingen, Germany
}

\date{February 2, 2012}

\begin{abstract}
  We solve the classical ``Game of Pure Strategy'' using linear
  programming. We notice an intricate even-odd behavior in the
  results of our computations, that seems to encourage odd or maximal
  bids. 
\end{abstract}

\maketitle

\section{Introduction}
Simple, two-player games are important models for human decision
making; they should have sufficiently elementary rules that they can
be studied both theoretically and empirically, yet be sufficiently
rich to involve a non-trivial amount of human psychological
experience.

We study, and solve numerically, the ``Game of Pure Strategy'', a
model of decision making based on bidding. It is an important paradigm
in game theory, because it can easily be shown (see below) that
\emph{no} deterministic strategy may succeed.

We notice, in the numerical data, that the parity of the bid is often
more important than its actual value. This echoes a known
recommendation for online bidding (in which the bidding amounts are
less restricted than in room bidding): the website
\verb+www.bidnapper.com+ recommends to its customers to ``Bid in odd
amounts. Many novices bid in rounded numbers.''

The computations were made with the help of the RRZN in Hannover,
whose support is gratefully acknowledged.

\section{The Game of Pure Strategy}
Goofspiel, also called Game of Pure Strategy (GOPS) is a two person
game.  Take a standard 52 card deck and discard all of the cards of
one suit.  The cards of one suit are given to one player, the cards of
another suit are given to the other player, and the cards in the
remaining suit are shuffled and placed face down in the middle.  The
cards are valued from low to high as ace=1, 2, 3, \ldots, 10, jack=11,
queen=12, and king=13.

A \emph{round} consists of turning up the next card from the middle
pile and then the two players `bet' on the upturned card, each player
choosing one card and then simultaneously displaying it to the other
player. The player showing the highest card wins the value of the
upturned card. If both players display the same card, the point value
is split between the two players. These three cards are then
discarded.  The game ends after 13 rounds and the winner is the person
who obtained the most points (you need 46 or better to win).

Though the mechanics of the game are simple, the strategy is not.
Suppose for example that the king is the upturned card in the first
round.  Further suppose that you choose to bet one (i.e. the ace).
When you turn your card up, you found out that your opponent bet his
king winning 13 points.  You are happy with this result because you
now have 12 more betting points which should more than make up for the
lost 13 points.  In fact, it is possible that you could win
\emph{every} remaining point by always betting one more than your
opponent (though of course that would require cheating, by knowing in
advance what your opponent is going to bet).  However, you are taking
a chance by betting only one: if your opponent had bet a two or three,
then she would have won 13 points at almost no cost.  When playing the
game you are trying to outguess your opponent while your opponent is
trying to outguess you: you find yourself reasoning along lines such
as ``my opponent is probably going to play $X$ so I should play $Y$,
but he may see through this and instead play $Z$ to defeat $Y$ so
maybe I should play $W$ instead.''

\section{Solving GOPS}
To make use of game theory, we use an equivalent scoring system: the
player with the higher card wins from the opponent the value of the
upturned card, or wins nothing in the event of a tie.  The game is
now a two-person zero-sum game that can be represented by a matrix
with one row for each possible play of player one and one column for
each possible play of player two.  The $i,j$'th entry of the matrix is
the value of the game to player one when player one makes his $i$'th
play while player two makes her $j$'th play (such a formulation is
called a \emph{matrix game}).

It is not hard to see that you should \emph{not} choose a
deterministic strategy. In fact, \emph{every} deterministic strategy
\textbf{A} can be defeated as follows.  Use strategy \textbf{A} to find the
card that my opponent is going to play. If my opponent is going to
play a King, play the ace.  Otherwise play the card that is one higher
than my opponent's choice.  This counterstrategy will win every round
except one resulting in a trouncing.  Instead your strategy should
have some random variations where you play particular cards with some
probability.

How difficult is it to analyze this game?  Suppose the cards are
valued from $1$ through $N$.  The number of distinct ways that the
middle suit could be ordered and the number of distinct betting
sequences for each player is $N$ factorial.  Hence, the number of
possible ways of playing out a game is $f(N) = (N!)^3$. Analysis of
GOPS, along these lines, would require consideration, for $N=13$, of
$2.4\times10^{29}$ variations, a number clearly beyond computational
possibilities.

To our knowledge, the game had never been previously analyzed beyond
$N=5$, see~\cite{Kerr}. There is a way to signficantly reduce the
number of games needed to be analyzed.  Sheldon Ross \cite{Ross}
describes a recursive rule expressing the value of a game as a
function of the values of smaller games. We give a further
simplification of his rule.

Let $f(V,Y,P)$ be the value for player one of the game in which $V$ is
the set of cards player one has in his hand, $Y$ is the set of cards
player two has in her hand, and $P$ is the set of cards in the
deck. Further, for $P_k\in P$, let $f_k(V,Y,P)$ be the value for
player one of that game, once the upcard $P_k$ has been revealed. Clearly
$f(V,Y,P)$ is the average of the $f_k(V,Y,P)$:
\[f(V,Y,P)=\frac1{|P|}\sum_{P_k\in P}f_k(V,Y,P).
\]
Suppose $|V| = |Y| = |P| = N$. Then $f_k(V,Y,P)$ is expressed as the
value of the $N \times N$ game whose payoff matrix $[X_{ij}]$ is
\[X_{ij} = P_{k} \operatorname{sign}(V_i-Y_j) + f(V\setminus\{V_{i}\},Y \setminus\{Y_{j}\},P\setminus\{P_{k}\});\]
here
\[\operatorname{sign}(x)=\begin{cases} 1 & \text{ if }x>0,\\
  0 & \text{ if }x=0,\\-1 & \text{ if }x<0.\end{cases}
\]

In english, this self-evident rule says the value of the game when
player one plays $V_{i}$ and player two plays $Y_{j}$ is the value of
the upturned card that is won or lost, plus the average value of the
remaining game where the average is taken over all possible remaining
upturned cards.

Blindly applying this rule results in a straightforward recursive
program; however, evaluation of $f(V,Y,P)$ on sets of cardinality $N$
requires $N^3$ evaluations of $f$ on sets of cardinality $N-1$,
leading again to the $(N!)^3$ complexity estimate.

To avoid this issue, we use a bottom-up approach storing the values
$f(V,Y,P)$ of the subgames as we go.  We use these stored values when
computing the larger subgames.  Using this technique, called
\emph{dynamic programming}, we compute the value of each subgame only
once.  For an initial $N \times N$ game, this reduces the number of
subgames we need to solve and store to $\sum_{j=0}^Nj\binom{N}{j}^3$,
a much more feasible number. Furthermore, we may use the symmetry
between players one and two to gain an extra factor of two, since
$f(V,Y,P)=-f(Y,V,P)$ and $f(V,V,P)=0$; and at each step of the
algorithm we only need to store in memory the values of $f(V,Y,P)$ for
a given value of $j$. On a large computer with 1TB core memory, the
game is then solvable up to $N=16$.

\subsection*{Linear programs}
If all values $f(V,Y,P)$ are known, it is then easy to compute the
optimal playing strategy. Let us say that the remaining cards are
$V,Y,P$ respectively for player one, player two and the deck, and that
$P_k\in P$ has been turned up. Recall the payoff matrix $[X_{ij}]$
from the previous subsection. The optimal strategy, for player one,
will take the form of an list of probabilities $x_i$ of playing card
$V_i$. Assuming that player two plays optimally, we want to maximize
$\min_j\sum_i x_i X_{ij}$; namely, we want to maximize the outcome,
allowing player two to make the best move (= minimize the outcome) as
long as she doesn't know our move. The solution is then a \emph{Nash
  equilibrium} of the game.

This maximization problem is a \emph{linear program} (LP), and we will
solve it using linear programming tools. The classical
reference~\cite{Dantzig} remains an excellent introduction to linear
programming. For example, suppose we have the following $3 \times 3$
matrix game:
\[
\left[
\begin{array}{rrr}
1 & -2 & 3 \\
-4 & 5 & -6 \\
7 & -8 & 9
\end{array}
\right].
\]

To formulate this as a LP, we introduce the variables $x_{1}, x_{2},$
and $x_{3}$ to represent the probabilities with which player one
should play columns 1, 2, and 3 respectively.  We also introduce the
variable $v$ to represent the value of the game.  The LP formulation
of this game is as follows:
\[
\text{maximize $v$ such that}
\left\{\begin{array}{rcrcrcccc}
x_{1} & - & 4 x_{2} & + & 7 x_{3} & - & v & \ge & 0,\\
-2 x_{1} & + & 5 x_{2} & - & 8 x_{3} & - & v & \ge & 0,\\
3 x_{1} & - & 6 x_{2} & + & 9 x_{3} & - & v & \ge & 0,\\
x_{1} & + & x_{2} & + & x_{3} & & & = & 1,\\
x_{1} & , & x_{2} & , & x_{3} & & & \ge & 0.
\end{array}\right.
\]

The last two constraint rows are needed to ensure that $x_{1}, x_{2},$
and $x_{3}$ form a probability distribution. The variable $v$ is
unrestricted.  Note that we are maximizing the expected profit, and
not the probability of winning.  If we are playing for money, and
there is some agreed-upon amount per point won, then this is the
optimal strategy. If, however, we want to maximize the
\emph{probability} of winning, and not the amount won, then the
results may be different.

Indeed, suppose the remaining cards are $queen,king$, player one has
$2,4$ in his hand, and player two has $ace,3$ in her hand. Player one
can guarantee victory by always playing $4$ when the king shows up,
but by doing so forfeits the chance of winning both last cards, so
guarantees a win by only one point. Playing either card with same
probability gives him an average gain of $12.5$ (the optimal strategy
is to play high with 52\% probability on the king, resulting in an
average gain of $12.52$).

There is a single $0 \times 0$ game, whose value is $0$, and it may be
used to start the induction with
$f(\emptyset,\emptyset,\emptyset)=0$. We note that, up to $2\times 2$
games, the results are easily computed by hand. Trivially, a
$1\times1$ game is won by the player having the largest card. Consider
the following $2 \times 2$ matrix game:
\[
\left[
\begin{matrix}
a & b \\
c & d
\end{matrix}
\right]
\]

If a value is a minimum value in its column and a maximum value in its row,
then it is a \emph{saddle point}.  If the game has a saddle point, then the
value of the game is the value of a saddle point entry (there may be more than
one saddle point).  If the game has no saddle point, then the value of the game
is $(ad-bc) / (a-b-c+d)$.  The formulation of a matrix game as a LP and the
solution to $2\times 2$ matrix games can be found in many sources (e.g.
\cite{Mendelson}).

\section{Results}
We have computed the winning strategies for $N=13$ using the method
described in the previous section: for each triple $V,Y,P$ of subsets
of $\{1,\dots,13\}$ of same cardinality, we have computed the value
$f(V,Y,P)$ of the corresponding game, and have computed the
probability arrays $x_{ki}$ with which card $V_i$ should be played if
the upcard is $P_k$.

Because of its formulation as a maximization of a piecewise-linear
function, these probabilities are rational numbers. We shall argue
that their denominators are so large as to make exact computations
pointless.

\subsection{Implementation}
We use the publicly available GLPK linear programming solver to solve
repeatedly the matrix games. This package implements the simplex
algorithm both in floating-point and multi-precision rational
arithmetic.

In our computer program, we represent the card sets by bit-vectors.
To conserve space, we use a \emph{perfect hash table}, that is a table
whose entries correspond bijectively to subgames. The subgames are
stored in lexicographic order; each subgame is represented by the
concatenation of the representations of player 1's hand, player 2's
hand, and the deck.  When we need the value of a previously computed
subgame, we compute its position in our ordering and grab the
corresponding entry from the table.  In our dynamic programming
method, we compute subgames in increasing size of hands. To conserve
space, we only store the results of the subgames of the current size
we are working on and the subgames of the next smaller size. This is
possible because the value of a subgame is needed only when computing
the values of subgames of the next larger size.

The results don't get interesting until $N=5$. There, in the initial
position, the optimal betting strategies, rounded to four digits, are

\begin{table}[ht]
\caption{Optimal strategy at the first move, $N = 5$}
\begin{tabular}{c|rrrrr|}
  & \multicolumn{5}{c}{\Large upcard} \\
  & {\bf 1} & {\bf 2} & {\bf 3} & {\bf 4} & {\bf 5} \\ \hline
  {\bf 1} & 0.0470 & 0.1855 & 0.1182 & 0.1226 & 0.1123 \\
  {\bf 2} & 0.8327 & 0 & 0.1188 & 0.07347 & 0.0241 \\
  {\bf 3} & 0.1203 & 0.7375 & 0 & 0.1915 & 0 \\
  {\bf 4} & 0 & 0.0770 & 0.7630 & 0.2043 & 0 \\
  {\bf 5} & 0 & 0 & 0 & 0.4081 & 0.8636 \\\hline
\end{tabular}
\end{table}

The exact values, as pointed before, are prohibitively long to write
down. For example, the top entry $0.0470$ is really
\begin{tabbing}
5306287082133275981487303358632452609704399851215456295628045583\\
1875882988814425504038999635791836938252991208618766205613347715\\
8328568084730206912059640079498714570709437787806964249018639429\\
44810525052774918079399954562997646185715598682274851480804900 /\\
1129268720669360758902051860285520214948806792442363910626460283\\
5249749724573405290487422550704989332508431906657764350941879381\\
7660750019182194563647083966290489533469158285998067618020193436\\
7037256681898791738381922918288920527729219085504200644992335857
\end{tabbing}

We have computed the exact values up to $N=7$; the numerators and
denominators of the optimal probabilities have approximately one
million digits.

\begin{table}[ht]
\caption{Optimal strategy at the first move, $N = 13$}
\small
\begin{tabular}{c|rrrrrrrrrrrrr|}
  & \multicolumn{12}{c}{\Large upcard} &\\
        &{\bf1}&{\bf2}&{\bf3}&{\bf4}&{\bf5}&{\bf6}&{\bf7}&{\bf8}&{\bf9}&{\bf10}&{\bf11}&{\bf12}&{\bf13}\\ \hline
{\bf 1} &    0 &    0 & .052 & .031 &    0 &    0 & .020 &    0 &    0 &    0 & .014 &    0 & .010 \\
{\bf 2} & .414 & .227 &    0 & .020 & .056 & .073 &    0 & .034 & .047 & .036 &    0 & .030 &    0 \\
{\bf 3} & .090 & .022 & .178 & .095 & .036 & .002 & .069 & .021 & .002 &    0 & .041 &    0 & .037 \\
{\bf 4} & .496 & .299 & .034 & .061 & .088 & .098 &    0 & .054 & .065 & .067 &    0 & .056 &    0 \\
{\bf 5} &    0 & .098 & .230 & .134 & .067 &    0 & .124 & .039 &    0 &    0 & .080 &    0 & .065 \\
{\bf 6} &    0 & .355 & .092 & .107 & .124 & .185 &    0 & .077 & .120 & .098 &    0 & .082 &    0 \\
{\bf 7} &    0 &    0 & .274 & .175 & .101 & .002 & .168 & .060 & .001 &    0 & .102 & .008 & .087 \\
{\bf 8} &    0 &    0 & .139 & .154 & .165 & .218 & .021 & .103 & .142 & .138 & .016 & .099 &    0 \\
{\bf 9} &    0 &    0 &    0 & .221 & .148 & .045 & .202 & .092 & .029 & .015 & .123 & .028 & .126 \\
{\bf 10}&    0 &    0 &    0 &    0 & .215 & .266 &    0 & .144 & .177 & .170 &    0 & .124 & .013 \\
{\bf 11}&    0 &    0 &    0 &    0 &    0 & .110 & .397 & .151 &    0 & .063 & .253 & .065 &    0 \\
{\bf 12}&    0 &    0 &    0 &    0 &    0 &    0 &    0 & .226 & .417 & .241 & .023 &    0 &    0 \\
{\bf 13}&    0 &    0 &    0 &    0 &    0 &    0 &    0 &    0 &    0 & .170 & .348 & .508 & .661 \\\hline
\end{tabular}
\end{table}

There are only a few general patterns in this table (and in the
corresponding ones for $N=6,\dots,12$ that we do not display). If the
upcard is large, say $>3N/4$, then it seems one should not bet a card
$<N/2$ and of opposite parity. Counterintuitively, sometimes one
should bet more than double the upcard. Also, when the initial card is
$N$, it seems that one should never bet $N-1$ nor $N-2$. However, it
is striking that, in the last column, odd bets are consistently (up to
$i=10$) to be preferred to even bets; and that, in general, one should
bet a card of the same parity as the upcard.

\section{Outlook}
The first author made a version of the program that stored the actual
probability vectors associated with the optimal strategies.  These
strategies were then used in a simple program that actually played the
9-card game.  Despite the counterintuitive nature of these results,
the computer player did win the majority of the games.  If you are
trying to maximize the probability of winning instead of amount won,
then there is at least one weakness in the computed strategies.
In the 9-card game, suppose the initial upcard is a 9.  The computer
player will play a 9 with a probability, rounded to four digits, of 0.7475.
Now if you always play a 1, then nearly 3/4 of the time you will gain an
advantage (playing a 1 against the computer's 9 is to your advantage
due to your increased betting strength for the remainder of the game)!
The computed strategies are optimizing the amount won not the probability
of winning.  When you play a 1 against the computed strategies, then in
a minority of cases the computer will play a small value winning the 9
and gaining more of an advantage than you stand to get when the computer
chooses to play a 9.  The optimal strategies for maximizing the
probability of winning are still unknown.

We have not been able to determine the exact value $\eta$ of an extra
bidding chip, though experimentation shows that, as $R$ gets large,
the values converge to the estimation $1+1/(2N-2)$, albeit very
slowly.

Our computer code was run on a 64-bit computer with 160 nodes and
640GB core memory. It relies on the GLPK linear programming
library. It is available, as a C++ program, at
\[\verb+http://www.uni-math.gwdg.de/laurent/gops.zip+\]


\begin{thebibliography}{1}

\bibitem[Dantzig (1963)]{Dantzig}
Dantzig, George B.
\newblock {\em Linear programming and extensions}
\newblock Princeton University Press, {Princeton, N.J.}, {1963}.
\newblock {xvi+625} pages.


\bibitem[Kerr (online)]{Kerr}
David Kerr.
\newblock http://home.netcom.com/\char126goldkerr/gops.htm

\bibitem[Mendelson (2004)]{Mendelson}
Elliott Mendelson.
\newblock {\em Introducing Game Theory and Its Applications}
\newblock Chapman \& Hall/CRC, Washington D.C., 2004.

\bibitem[Ross (1971)]{Ross}
Sheldon M. Ross.
\newblock Goofspiel: The Game of Pure Strategy.
\newblock {\em Journal of Applied Probability,} Vol.8, No.3 (Sept.71), pp.621-25.

\end{thebibliography}
\end{document}